\documentclass[12pt,preprint]{aastex}

\usepackage{graphicx}
\usepackage{natbib}
\usepackage{amssymb}
\usepackage[figuresright]{rotating}

\slugcomment{To appear in Ap. J.}

\shorttitle{Spitzer-MIPS survey of VMR-D} \shortauthors{Giannini
et al.}

\begin{document}


\title{Spitzer-MIPS survey of the young stellar content in the 
Vela Molecular Cloud-D}


\author{T.Giannini\altaffilmark{1}, D.Lorenzetti\altaffilmark{1}, 
M. De Luca\altaffilmark{1,2}, B.Nisini\altaffilmark{1}, 
M.Marengo\altaffilmark{3}, L.Allen\altaffilmark{3}, H.A.Smith\altaffilmark{3},
G.Fazio\altaffilmark{3}, F.Massi\altaffilmark{4}, D.Elia\altaffilmark{5}, 
F.Strafella\altaffilmark{5} \\
}
\altaffiltext{1}{INAF - Osservatorio Astronomico di Roma, via Frascati 
33, 00040 Monte Porzio, Italy, giannini, deluca, dloren, 
nisini@oa-roma.inaf.it}
\altaffiltext{2}{Dipartimento di Fisica - Universit\`a di Roma `Tor Vergata',
via della Ricerca Scientifica 1, 00133 Roma, Italy}
\altaffiltext{3}{Harvard-Smithsonian Center for Astrophysics, Cambridge, MA,
USA}
\altaffiltext{4}{INAF - Osservatorio Astrofisico di Arcetri, Largo E. Fermi 5,
50125 Firenze, Italy}
\altaffiltext{5}{Dipartimento di Fisica - Universit\`a del Salento, CP 193, 73100,
Lecce, Italy}



\begin{abstract}
A new, unbiased Spitzer-MIPS imaging survey ($\sim$ 1.8 square degrees) 
of the young stellar content of the Vela Molecular Cloud-D is presented. 
The survey is complete down to 5 mJy and 250 mJy at 24~$\mu$m
and 70~$\mu$m, respectively. A total of 849 sources are detected at 24~$\mu$m
and 52 of them also have a 70~$\mu$m counterpart. The VMR-D region is one that we have already partially mapped in dust
and gas millimeter emission, and we discuss the correlation between the Spitzer compact sources
and the mm contours.  About half of the 24\,$\mu$m sources are located inside the region 
delimited by the $^{12}$CO(1-0) contours, corresponding to only  one third of the full area 
mapped with MIPS.  Therefore the 24\,$\mu$m source density increases by about 100$\%$
moving from outside to inside the CO contours. For the 70~$\mu$m sources, the
corresponding density increase is four times. About 400 sources of these have a 2MASS
counterpart, and we have used this to construct a K$_s$ vs. K$_s$-[24] diagram 
and to identify the protostellar population inside the cloud. 

We find an excess of Class~I sources in VMR-D in comparison with other star forming regions.
This result  is reasonably biased by the sensitivity limits at 2.2 and 24~$\mu$m, 
or, alternatively, may reflect a very short
lifetime ($\la$10$^6$~yr) of the protostellar content in this molecular cloud. The MIPS images
have identified embedded cool objects in most of the previously identified starless cores in the
region; in addition, there are 6 very young, possibly Class 0 objects identified. Finally we report
finding of the driving sources for a set of five out of six very compact protostellar jets that had been
previously discovered in near-infrared images of VMR-D. 
\end{abstract}

 

\section{Introduction}

Infrared maps of star forming Giant Molecular Clouds (GMCs) are an essential tool in the
modern study of star formation.  When radio and millimeter maps also exist, the relationships
between the regions of infrared, millimeter and radio activity provide one of the key new tools
for clarifying the varieties of star formation that can occur. The sensitivity of infrared techniques
means that even shallow surveys can in principle reveal the processes of both low and high mass
star formation in clouds that are not too far away.   There are,  however, very few nearby GMCs
with which to take full advantage of these techniques. Among these available targets, one is the
Vela Molecular Ridge (VMR), a complex of four adjoining GMCs (Murphy \& May 1991;
Yamaguchi et al. 1999), located in the galactic plane ($b$=$\pm$3$^{\circ}$) outside the solar
circle ($l \sim$ 260$^{\circ}$ - 275$^{\circ}$);  most of the gas (clouds named A, C and D) is
located at a distance of about 700 pc (Liseau et al. 1992).

This team has studied the star formation activity in the VMR for many years: the concentration of
red and young sources (Liseau et al. 1992, Lorenzetti et al. 1993); the presence of embedded
clusters (Massi et al. 2000, 2003); the occurrence of protostellar jets (Lorenzetti et al. 2002;
Giannini et al. 2001, 2005, De Luca et al. 2007, hereinafter D07). Recently we mapped with the SIMBA
bolometer array at SEST a $\sim$ 1 deg$^2$ area of the cloud D in the 1.2\,mm continuum of dust emission, 
and in the $^{12}$CO(1--0) and $^{13}$CO(2--1) transitions (Massi et al. 2007, hereinafter M07; Elia et
al. 2007, hereinafter E07).

The advent of the Spitzer Space Telescope (SST, Werner et al. 2004) and the imaging
photometric facilities on board, i.e. the Multiband Imaging Photometer for Spitzer (MIPS, 24, 70, 
160\,$\mu$m; Rieke et al. 2004) and the InfraRed Array Camera (IRAC, Fazio et al.
2004)   has enabled us to obtain maps of the VMR from 3.8 to 70~$\mu$m across the same area
already surveyed in the millimeter emission of dust and gas. The primary goal of this survey is to
obtain a census of the embedded young stellar population of VMR-D and to correlate it with
its gas and dust cores. 

This paper describes our MIPS observations of the VMR-D; it is the first of a series of papers we
are preparing dealing with VMR-D as seen by Spitzer; the IRAC data of the same region will be
presented in a separate paper. Spitzer surveys of several other star
forming regions have been already published, most of them in the framework of the cores-to-disk
(c2d) legacy project (Evans et al. 2003). Most of these regions, however, are located outside the
Galactic plane ($|b|$ $>$ 10$^{\circ}$) in regions that were originally selected in part to avoid
strong confusion and extinction problems. A huge amount of observational material has been so
far accumulated and published on those clouds. For the VMR, our current multi-frequency
database, when combined with the increased sensitivity of current instruments, has allowed us to
overcome many of the problems associated with observations of GMCs in the galactic plane. 
Since the plane is  where most of the material currently forming stars is located, it is both a
natural and critical region to understand, and will also help with the comparison between the
derived properties of our Galaxy with those of external galaxies, whose planes are the unique
zones we are able to sample.

Spitzer has surveyed many different types of star formation regions.  
Therefore, legitimate comparisons between them all would benefit from a standard analysis and
presentation, although some problems could arise because their numerous differing parameters, 
as well as the various details of the observations. In this paper we therefore adopt as much 
as possible the methods that have already been used successfully in the c2d program. 

Our paper is organized as follows: in Sect.\,2 we give the details of the observations and data
reduction procedure; the results are presented in Sect.\,3 and discussed in Sect.\,4 and 5.
Concluding remarks are given in Sect.6.

\section{Observations} 

VMR-D cloud was observed with MIPS on board the Spitzer Space Telescope
within the Guaranteed Time Observation program (PID 30335).
The observations covered $\sim$ 1.15 (in R.A.) $\times$ 1.6 (in dec.) degrees centered 
at $\alpha$, $\delta$ (J2000) = 8$^h$~47$^m$~50$^s$,
-43$^{\circ}$~42$^{\prime}$~13$^{\prime\prime}$ 
and 8$^h$~48$^m$~20$^s$, -43$^{\circ}$~31$^{\prime}$~26$^{\prime\prime}$  at 
24~$\mu$m and 70~$\mu$m, respectively (orientation: 145$^{\circ}$ W of N).

Data were collected on 14 Jun 2006 in scan mode, medium speed, with 5 scan legs 
and 160$^{\prime\prime}$ cross-scan step, resulting in a total integration time 
of 40 seconds (for both 24 and 70~$\mu$m) per pixel.
The mapping parameters were optimized for the 24 and 70~$\mu$m bands: as a consequence, 
the 160~$\mu$m map suffers from coverage gaps and saturation and will be not considered 
in the following.

The SSC-pipeline, version S14.4.0, produced basic calibrated data (BCDs) 
that we have used to obtain mosaiced, pointing refined images by means of 
the MOPEX package provided by the Spitzer Science Center (Makovoz \& Marleau 2005).

The main instrumental artifacts have been removed from the mosaiced images 
by means of the MOPEX package. Minor problems of residuals jailbars 
(expecially at 70~$\mu$m) and background matching between adjacent frames 
(at 24~$\mu$m) are still visible close to the brightest objects, but they do 
not affect significantly the point source photometry discussed in this paper.

The final 24\,$\mu$m map global properties can be summarized as follows: 
pixel scale of 2.45$^{\prime\prime}$/pixel, background r.m.s. of 0.3~$\mu$Jy/arcsec$^2$, 
within the regions of high level of diffuse emission. The brightest sources saturate at the
emission peak: for these we estimate a lower limit to the integrated flux of 4 Jy.
In the 70\,$\mu$m map the pixel scale is 4.0\,$^{\prime\prime}$/pixel and the background r.m.s.
ranges between 23 and 94\,$\mu$Jy/arcsec$^2$. None of the detected sources appears saturated
at this wavelength.

\section{Results} 

\begin{figure*}
\includegraphics[angle=0,scale=0.8]{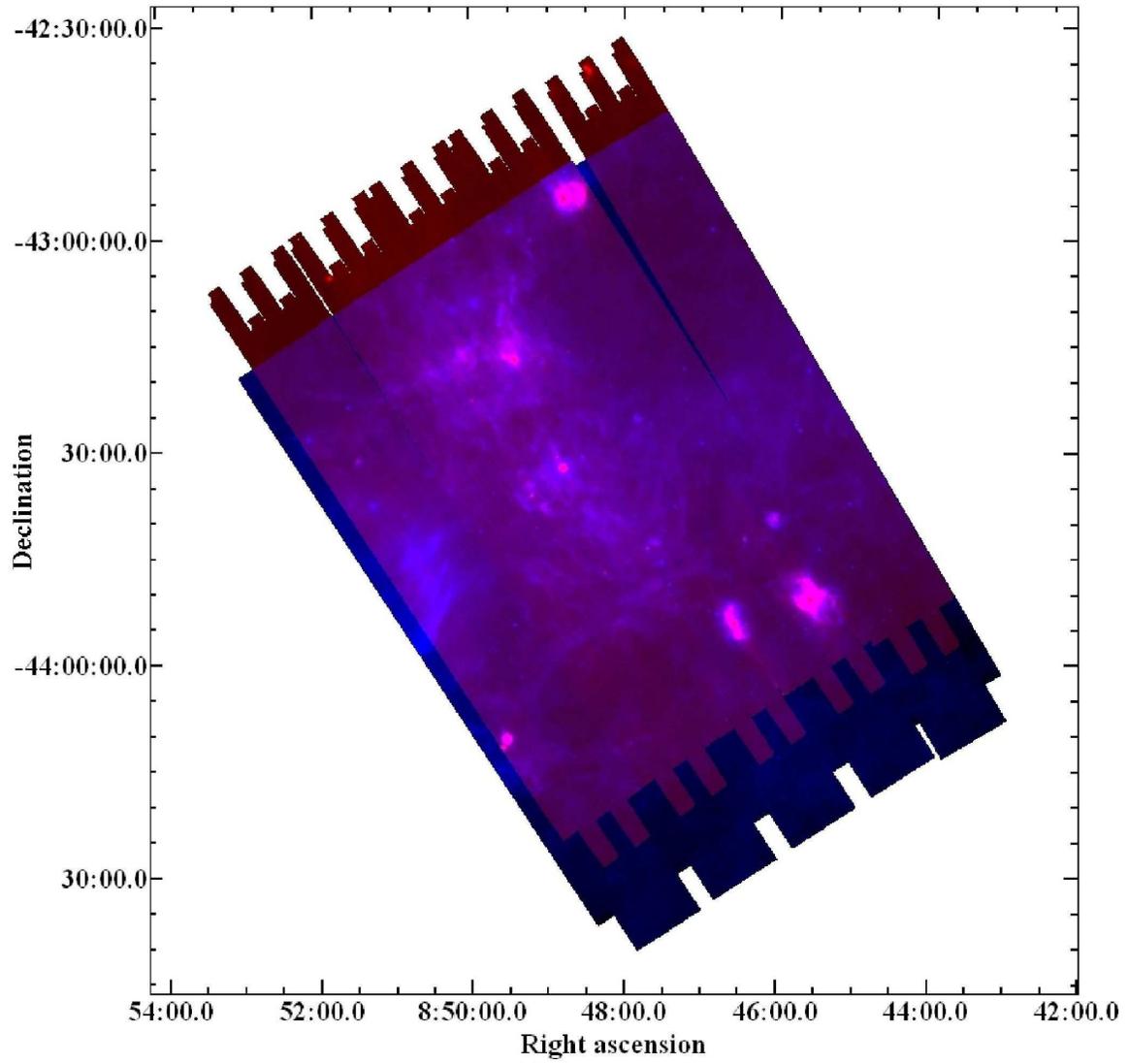}
\caption{MIPS two-color map (24\,$\mu$m in blue, 70\,$\mu$m in red) of
VMR-D.
\label{MIPS24-70}}
\end{figure*}

\begin{figure*}
\includegraphics[angle=0,scale=0.8]{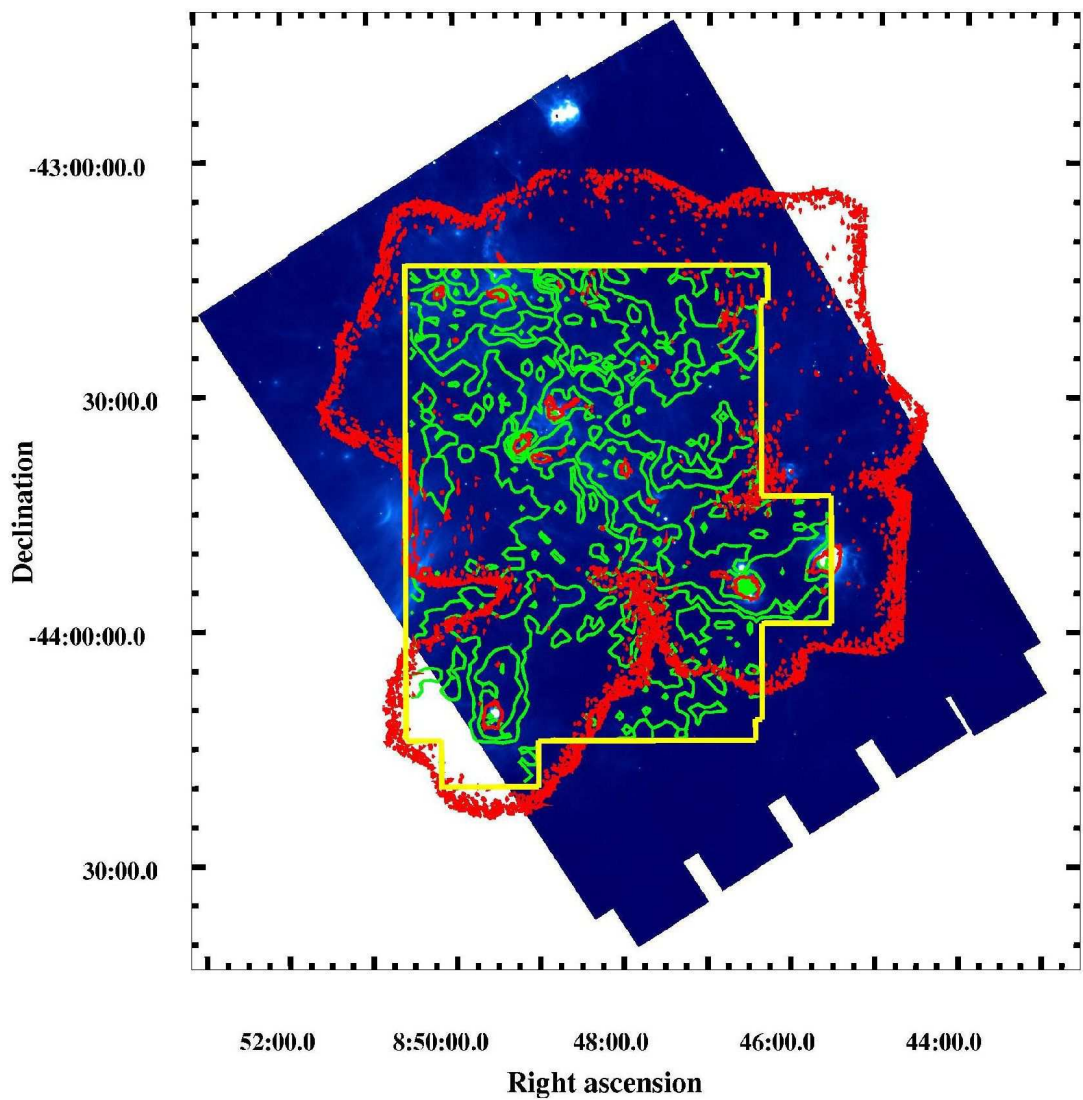}
\caption{Mosaic of VMR-D map at 24~$\mu$m, with superposed 
the $^{12}$CO intensity map (whose limits are depicted in yellow), where the contours (in green)
are in the range -2\,-\,20 km s$^{-1}$
(adapted by E07). Also overlaid is the 1.2$\it{mm}$
dust emission map (red contours, adapted from M07).
CO contour levels start from 5 K km s$^{-1}$ and are in steps of
25 K km s$^{-1}$, while dust contours start from 50 mJy/beam and
are in steps of 50 mJy/beam.  \label{MIPS24}}
\end{figure*}

\begin{figure*}
\includegraphics[angle=0,scale=0.8]{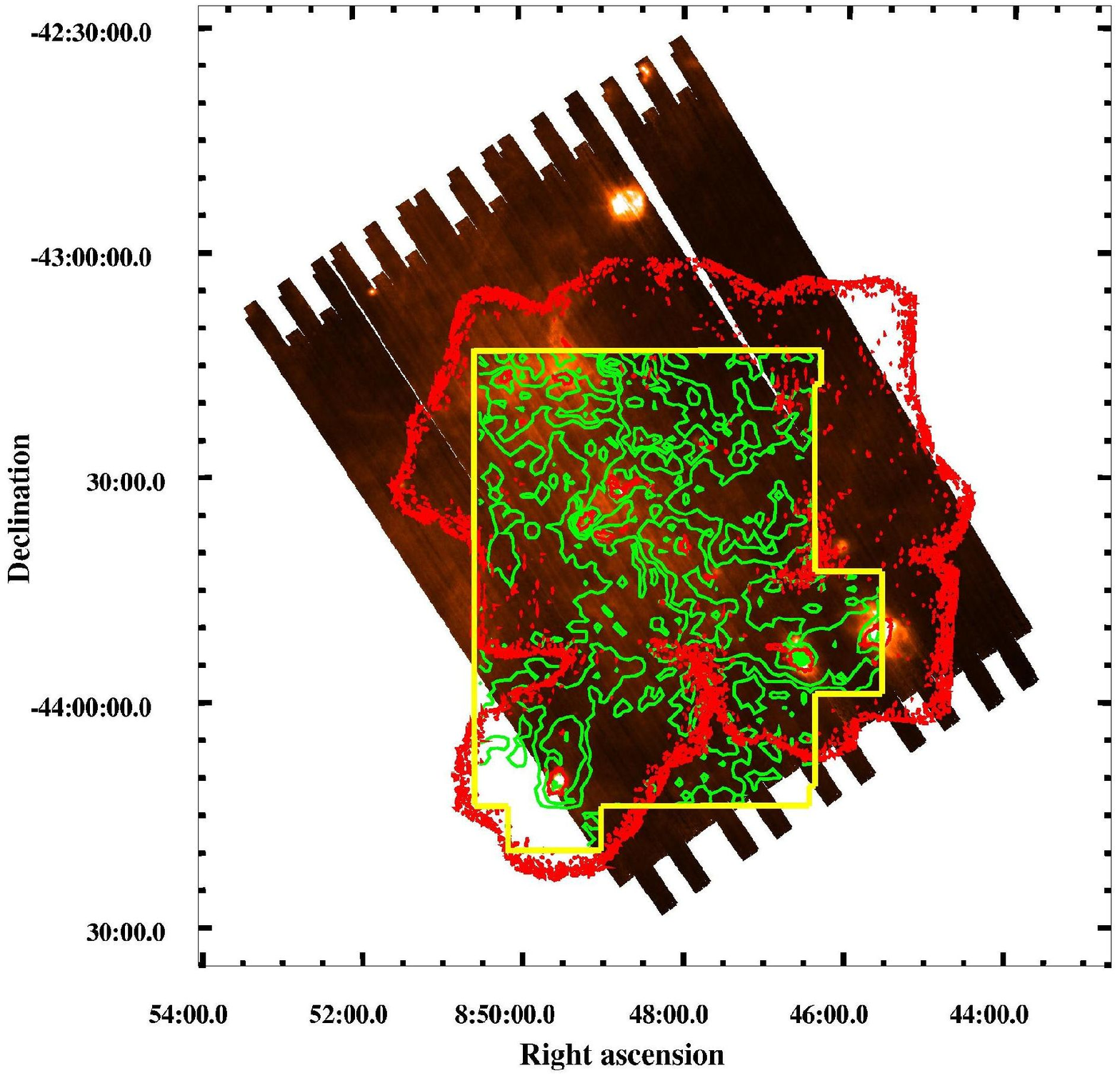}
\caption{As Figure~\ref{MIPS24} for the 70\,$\mu$m map.\label{MIPS70}}
\end{figure*}

Figure\,\ref{MIPS24-70} shows the two-color final mosaic of VMR-D
(24\,$\mu$m in blue, 70\,$\mu$m in red), while on Figures \ref{MIPS24} and 
\ref{MIPS70}, images in each of the two filters are shown separately.
In these latter, the 1.2\,mm dust map (adapted from M07) and the $^{12}$CO intensity map
integrated in the velocity range -2\,$\div$\,20 km s$^{-1}$ (adapted from E07) 
have been superimposed for comparison.  We define as `on-cloud' all objects 
inside these latter contours. Such a definition is perforce just the first level effort of delimiting
the sources belonging to the molecular cloud; indeed, it is clear from Figures \ref{MIPS24} and
\ref{MIPS70} that 
the $^{12}$CO emission remains well above the 3$\sigma$ level at the north and west borders 
of the gas map, and thus sources belonging to VMR-D could exist toward these directions. 
Considering such sources as 'off-cloud' will have the effect of reducing the distinctions between
the 'on' and 'off' cloud populations; these sources should then be considered on a 
case-by-case basis (see sect.5.1). We have also considered as 'off-cloud' those regions where the CO peak velocity
is faster than  20~km s$^{-1}$, since they are likely to be more distant and unassociated
with VMR-D (see Figure~1 in Lorenzetti et al. 1993).

The point-source extraction and photometry processes were performed by using
the {\it DAOPHOT} task of the astronomical data analysis package {\it
IRAF}\footnote{IRAF, the Image Reduction and Analysis Facility, is a general
purpose software written and supported by the IRAF programming group at
the National Optical Astronomy Observatories (NOAO) in Tucson, Arizona 
(http://iraf.noao.edu).}.
Given the size of the MIPS mosaic it was impossible to apply any 
automatic procedure for finding sources down to the sensitivity 
limits without being affected by a locally varying background level; we 
therefore applied a searching algorithm as deep as possible, 
but still compatible with an automatic procedure. The search algorithm
was applied to a differential image we produced between the final mosaic and a 'sky' image, 
the latter obtained by applying to the mosaic a median filter over boxes of 5$\times$5 pixels.
A threshold of 30\,$\sigma$ has been imposed on the sky-subtracted image, which
corresponds at least to 5\,$\sigma$ (depending on the local background) 
in the unsubtracted image.

\begin{figure}
\includegraphics[angle=-90,scale=.40]{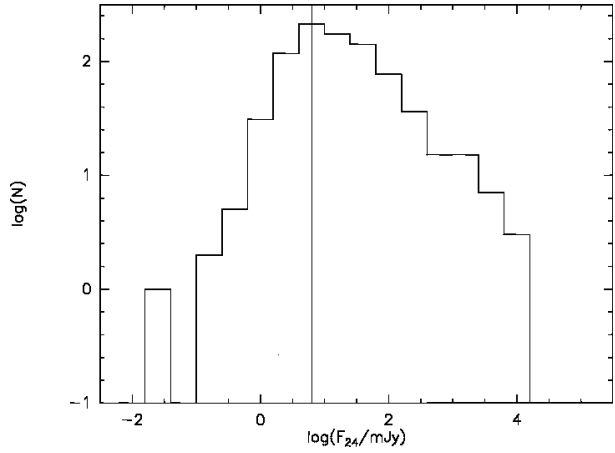}
\caption{Histogram of the sources detected at 24~$\mu$m. The completeness
limit is around 5 mJy, as indicated by the vertical line.\label{compl24}}
\end{figure}

\begin{figure}
\includegraphics[angle=-90,scale=.40]{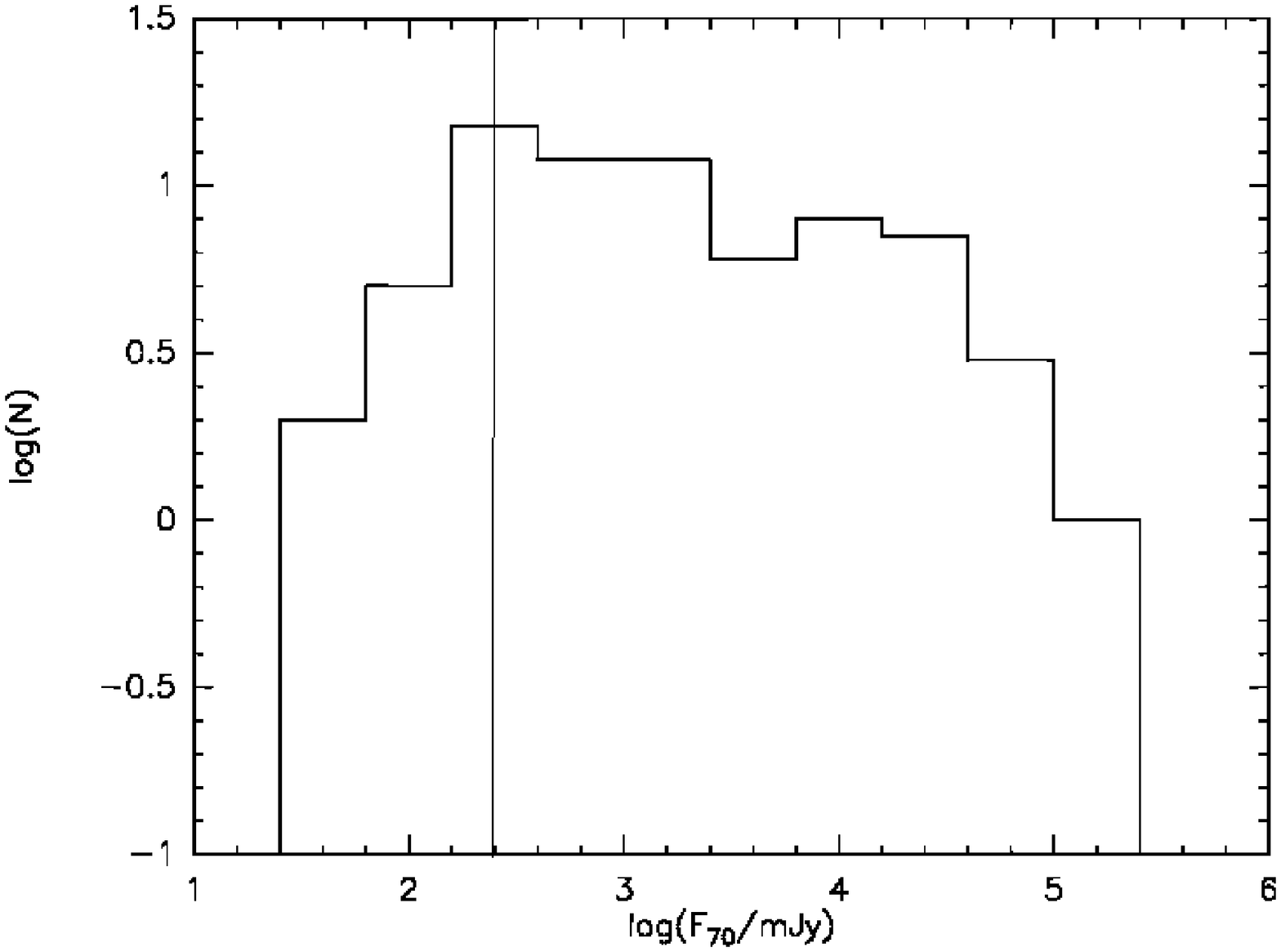}
\caption{Histogram of the sources detected at 70~$\mu$m. The completeness
limit is around 250 mJy, as indicated by the vertical line.\label{compl70}}
\end{figure}

\begin{figure*}
\includegraphics[angle=0,scale=1]{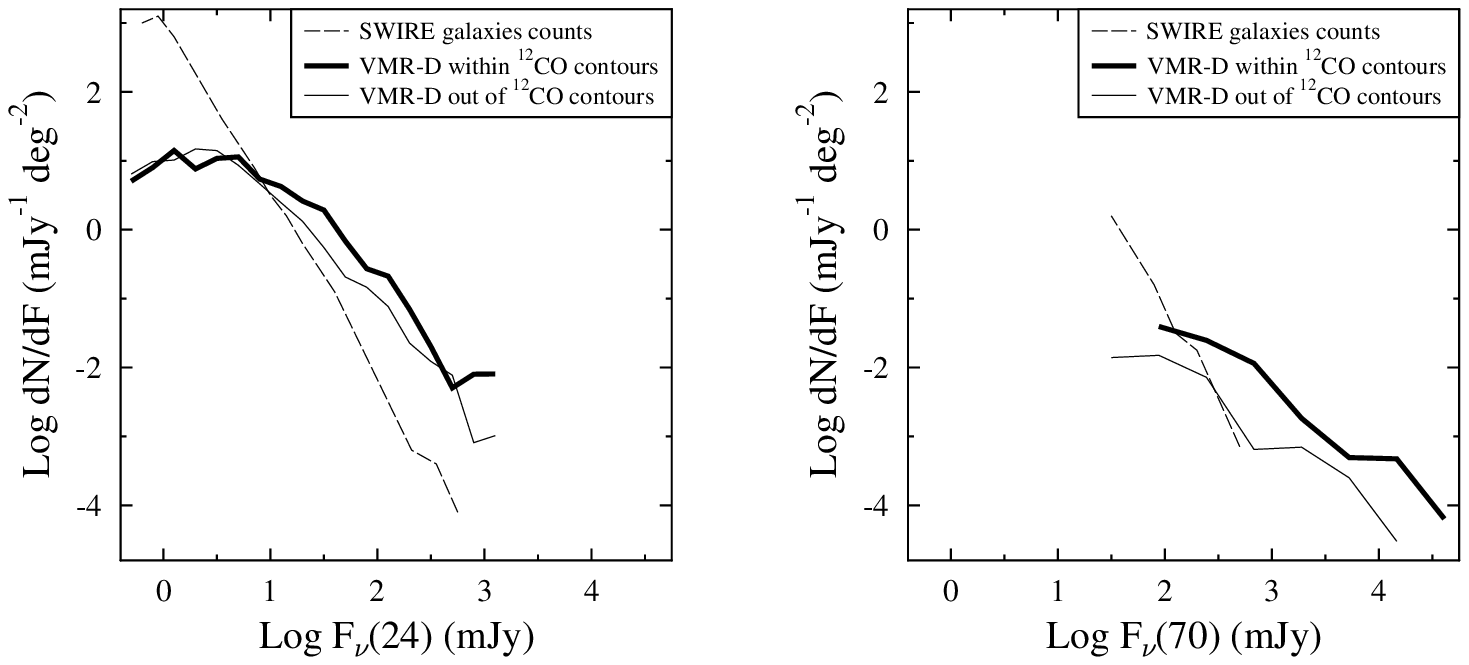}
\caption{Left panel: differential number counts at 24~$\mu$m. Thick and thin lines refer to
source in VMR-D within and outside the $^{12}$CO contours, respectively.
Extragalactic background sources from the SWIRE ELAIS N1 field are shown for 
comparison (these latter have been taken from Figs.6 and 7 in Rebull et al. 2007, hereinafter
R07). Right panel: as 
in left panel at  70~$\mu$m.  
\label{fig:counts24_70}}
\end{figure*}

The automated methods just described lead to the detection of 838 and 61 point sources 
at 24~$\mu$m and 70~$\mu$m, respectively. A further 12 detections have been added
to the 24~$\mu$m list by applying local sky values in selected areas (see the discussion below).
The source distribution as a function of the measured flux is depicted in Figures~\ref{compl24}
and \ref{compl70}, where the completeness limits can be evaluated as the flux bin corresponding
to the maximum counts before the decline at lower fluxes due to
the instrumental sensitivity.  We determine that our sample is complete down to 5 and 250 mJy at
24 and 70~$\mu$m, respectively. 

A statistical summary of the detected sources is presented in Table \ref{tab:tab1}. 
About 45\% of the 24~$\mu$m sources are spatially located inside the region 
delimited  by the $^{12}$CO contours ($\sim$ 0.61 deg$^2$), even though this latter 
is about one half the size of the remaining mapped area ($\sim$ 1.23 deg$^2$). 
This result gives an initial indication of how the IR source density increases by 
about 100\% moving from outside to inside the CO contours,  and the pattern becomes even
more significant when considering the 70\,$\mu$m sources, whose source density increase
is four times. 
The 24~$\mu$m counts per deg$^2$ are 
represented in Figure~\ref{fig:counts24_70}, left panel, where different symbols indicate those
sources located respectively within and outside the CO contours.
Also in this plot, where the differential number density is shown, there is a drop 
for F$_{\nu}$ $<$ 35\,mJy. For greater F$_{\nu}$  values the number of objects inside the gas
contours (i.e. those more likely associated to the cloud) systematically
exceeds the number of objects outside the cloud, giving  reasonable support to the
empirical significance of this crude classification. In addition,   Figure~\ref{fig:counts24_70}  is
a comparison between the on- and off-cloud samples and the Spitzer Wide-area Infrared
Extragalactic Survey (SWIRE, Lonsdale et al. 2003) legacy program. A
significant amount of contamination from the extragalactic background is predicted (at
24~$\mu$m) for flux densities $<$10 mJy down to the completeness limit, so that 'on' and 'off'
source populations at this level become undistinguishable. 

The counts per deg$^2$ at 70~$\mu$m are depicted 
in Figure~\ref{fig:counts24_70}, right panel: here again the possible extragalactic contamination
appears just at (or even below) the completeness limit.\\
In the same Figure~\ref{fig:counts24_70}, we also note that the 70~$\mu$m counts confirm the
on- and off-cloud distributions already found at 24\,$\mu$m.

The complete catalog of the detected sources is given in electronic form (a short sample version is printed in Table~\ref{tab:tab2}). In Table~\ref{tab:tab3}
we show the list of the 70~$\mu$m detections: of the  61 sources, 52 of them are 
coincident with a 24~$\mu$m source (i.e. the distance in both right ascension 
and declination is less than the 20$^{\prime\prime}$ PSF radius 
at 70 $\mu$m, see the summary of Table~\ref{tab:tab1}). In Table~\ref{tab:tab3}, we list the 
24~$\mu$m coordinates (which are more accurate than the 70~$\mu$m ones because of the
smaller PSF at 24~$\mu$m), the distance from the 70~$\mu$m coordinates, 
($\Delta\alpha$/$\Delta\delta$)$_{70}$, the measured flux at 24 and 70~$\mu$m along with 
the relative uncertainties, a flag indicating whether or not 
the source is located inside the region delimited by CO emission contours, and
the association with a dust core, if any. This latter is based on the distance between the 24$\mu$m
and mm coordinates, ($\Delta\alpha$/$\Delta\delta$)$_{mm}$, which must be within the SIMBA HPBW 
of 24$^{\prime\prime}$. All the dust cores associated with a 24~$\mu$m source are also associated with
its 70~$\mu$m counterpart.\\
Nine 70~$\mu$m sources have no 24~$\mu$m
counterpart.  Four of these were not imaged at 24~$\mu$m because of the shift between the two
maps, four appear as diffuse or with a filamentary structure at 24~$\mu$m, and one 
has F$_{24}$ $<$ 1.2 mJy (3$\sigma$ upper limit): this source (\#1 in Table~\ref{tab:tab3})
could be (if not a galaxy) a very young protostar which deserves further attention. 

We also provide in Table~\ref{tab:tab3} the association of
MIPS 24/70~$\mu$m sources with dust cores found in VMR-D by M07.
 The detailed study of these sources will be addressed in a future paper;  here we give some
preliminary results and point out some statistical aspects. In the region mapped in the dust
emission at 1.2\,mm 
(see Figures\,\ref{MIPS24} and \ref{MIPS70}), a robust sample of 29 cores has been revealed,
along with 26
cores whose size is below the map spatial resolution (24$^{\prime\prime}$). 
D07 have associated 12 of these cores 
(8 resolved and 4 under-resolved) with an IRAS or MSX point source, while the remaining 43
cores
are not associated with any FIR counterpart, so that they appear to be either cold Class
0 sources/starless cores (in case of resolved cores) or possibly data artifacts (in case of 
under-resolved cores). As stated in D07,  such a high fraction of starless cores as compared to
protostellar cores is most likely a result of the poor sensitivity of the IRAS/MSX facilities. Our
significantly more sensitive MIPS data offers the opportunity to check whether 
or not such a bias exists, and to eventually find weak counterparts of the dust cores.
In order to resolve this issue we closely reexamined our maps, performing photometry on the mm
peaks coordinates using local rather than global thresholds for the background level.  This
technique turned up 12 new objects at a flux density as low as 0.7 mJy at 24~$\mu$m, fainter
than the completeness limit by more than a factor of 7.

This procedure, together with automatic finding described above, 
when applied overall led to the association of 23 resolved and 20 under-resolved cores with 58
sources at 24 ~$\mu$m, 19 sources at 70~$\mu$m  (in some cases we found multiple
associations), thereby dramatically increasing the
percentage of cores associated with an embedded protostar from 22\% (D07) to 78\%. This
result is in general agreement with recent MIPS findings in other GMCs that have substantially
modified the percentage of active vs. inactive cores in favor of the former (e.g. Young et al.
2004).  We also note that the existence of a MIPS counterpart to 20 out of 26 under-resolved
cores significantly reduces the possibility that these objects are simply data artifacts.  The lack,
even at the MIPS sensitivity,  of a FIR counterpart to five
resolved dust peaks (namely MMS 6,\,13,\,15,\,20,\,24 in the list by M07) makes these objects  a
very robust sample of genuine starless cores.

\section{Comparison with IRAS sources}
The similarity of the MIPS 24 and 70~$\mu$m bandpasses to the 
25 and 60\,$\mu$m filters on-board IRAS offers us the opportunity to 
evaluate directly the reliability of the IRAS point source catalogue (IRAS-PSC) 
fluxes in crowded and diffuse clouds like VMR-D, objects that are commonly found 
in the galactic plane. A similar study has already been performed 
by R07 in the Perseus molecular cloud; although in this case
the geographic location makes extended emission and source 
confusion less critical, only 61\% (at 25$\mu$m) and 32\% (at 60$\mu$m)
of the objects of the IRAS-PSC are recovered by MIPS as point-like sources, while all
the others, although detected, remain confused by nebulosity. Higher rates of coincidence 
are found, at least at 25\,$\mu$m, if the Faint Source Catalogue (FSC) -
produced by point-source filtering the individual detector data streams - is used.

Unfortunately, the FSC does not
cover the galactic plane, so that we cannot confirm this result on VMR-D. 
Here, a total of 57 high (f$_{qual}$=3) or moderate (f$_{qual}$=2) quality 
detections are listed in the IRAS-PSC catalogue at 25\,$\mu$m;  46 of them (80\%)
are also seen by MIPS and recovered with our algorithm, while the remaining  11 IRAS objects
appear as diffuse 
emission at 24\,$\mu$m and are thus undetected as point-sources. 
The matching rate for VMR-D is thus higher than in Perseus.  The same trend is seen 
at 60\,$\mu$m, where out of 48 IRAS-PSC items, the retrieval rate is of about 50\%. 
In Table~\ref{tab:tab4} we give the list of the IRAS-PSC (with any f$_{qual}$) not 
recovered by MIPS. 
The IRAS sources in the table marked as `off-edge' in one MIPS bandpass are necessarily 
`on-edge' in the other, because of the spatial shift between the two focal plane
arrays.  Along with the f$_{qual}$ flag, we also give in Table~\ref{tab:tab4}
the IRAS correlation coefficient flag (cc) which provides an indication of the 
point-likeness confidence of the detected source. This flag is coded as alphabetical
character and subsequent letters correspond to decreasing accuracy (i.e. A$>$99\%,
E$>$96\%). 
Noticeably, PSC sources retrieved by MIPS (with f$_{qual}$=2,3), show, on average, 
`cc' flag equal to "A" or "B": such an occurrence can thus be translated into a
suitable tool to broadly distinguish between genuine point-source and diffuse emissions, if 
MIPS maps (and FSC detections) are unavailable.

\section{Color-Magnitude diagrams}

\subsection{K$_s$ vs. K$_s$-[24]}

\begin{figure}
\includegraphics[angle=0,scale=.50]{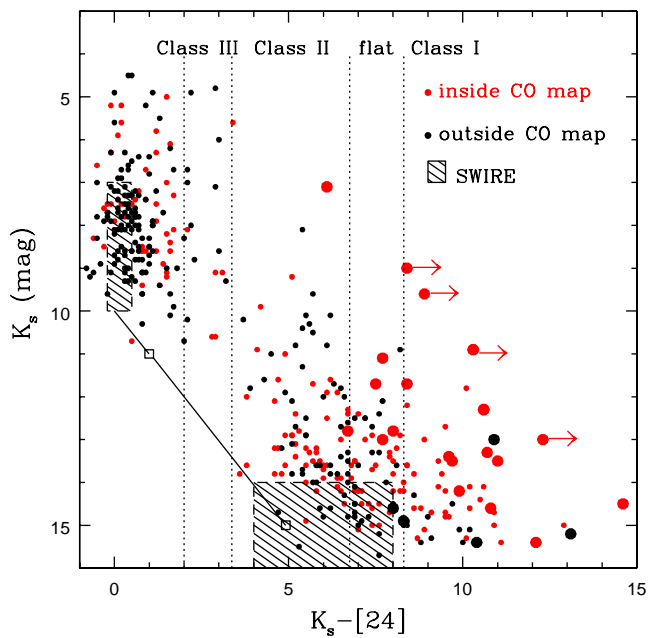}
\caption{Color-magnitude diagram for the 2MASS K$_s$-band
and the MIPS 24\,$\mu$m sources. Of the 849 24~$\mu$m sources 
in the MIPS map, 401 have a K$_s$ detection within a radius of 5 arcsec.
These are shown by red dots if located inside the CO contour map (180 sources)
and by black dots if outside (221 sources).
Large dots denote sources with 70~$\mu$m detections, while arrows refer to
sources saturated at 24\,$\mu$m. Hatched
areas are the {\it loci} of the sources in the SWIRE survey (taken from R07). 
The thick line indicates the effect of the extinction for different values of A$_V$ (open squares
refer to A$_V$=10 and 50~mag).\label{fig:k-24}}
\end{figure}

About half of the 24~$\mu$m detections have identifiable 2MASS counterparts 
at K$_s$ (limiting magnitude of 15.3) within a radius of 5$^{\prime\prime}$. 
These 2MASS fluxes have been used to construct the K$_s$ vs. K$_s$-[24] 
color-magnitude diagram given in Figure~\ref{fig:k-24}, where MIPS 
sources inside and outside the CO contours are shown with different colors. 
Also reported as hatched areas are the {\it loci} of the extragalactic sources in the SWIRE survey. 
As expected for a molecular cloud in the galactic plane, there are very few 
extragalactic sources seen. A remarkable number of objects fall at K$_s$$<$8.5 mag and K$_s$-[24]$\sim$0, 
which, given our completeness limit at 24~$\mu$m of 5~mJy, delimits the region
of normal photospheres in VMR-D.   
Noticeably, in this part of the diagram, the number density (per deg$^2$)
of the 'off-cloud' sources is larger than that of the 'on-cloud'
ones (100 vs. 69, see Table~\ref{tab:tab4}): in principle, all the unreddened
photospheres detected in VMR-D could be indeed foreground/background stars. More reasonably,
we can affirm that no increase of main-sequence stars (with respect to the
adjacent field) is registered in VMR-D, as expected because of the youth of the region.

The thick squares indicate the effects of an extinction of A$_V$ =
10 and 50\,mag, respectively, on the data.
The quantitative A$_V$ map of the overall region by Dobashi et al. (2005) does not 
provide values in excess to 5-10\,mag (below the saturation limit of the catalog of 15
mag), while toward the dust cores A$_V$ can increase up to
$\sim$ 20\,mag (M07 and E07).  We thus conclude that sources 
with K$_s$-[24] $>$ 5 are probably reasonably young objects, and not merely extinguished
objects. 
Indeed, it is not accidental that the large majority of these sources
belong to the molecular cloud, nor that the objects with normal photospheres are outside the
cloud.\\
From an evolutionary point of view, protostars can be characterized on the basis of the spectral 
classification between 2\,-\,10\,$\mu$m  (Greene et al. 1994), according to which
different evolutionary stages, from the accretion phase (Class I) to the beginning of 
the main-sequence (Class III), are manifested. 
The same authors have found that the 2\,-10\,$\mu$m spectral index  
does not change substantially when computed using fluxes up to 20\,$\mu$m 
(by using photometry in the Q band);  this result allows one to extrapolate the 24\,$\mu$m 
flux for different spectral indexes and accordingly to compute 
the expected value of the K$_s$-[24] color. 
The result of this procedure is given in R07, who furthermore
requested that, to select Class III sources from  
normal photospheres and foreground/background stars, K$_s$-[24]$>$2. The spectral classification
derived for VMR-D is depicted in Fig.\,\ref{fig:k-24} and also reported in Table \ref{tab:tab5}.
The ratio of `on-cloud' over `off-cloud' objects within the same class
increases with increasing K$_s$-[24] values; 
moreover, most of the younger 'on-cloud' objects are also detected in the 70~$\mu$m band
(large, red dots in Fig.\,\ref{fig:k-24}).
Noticeably, 70\% of the `off-cloud' objects showing the characteristics of the youngest
and coldest sources (black dots at K$_s$-[24]$\simeq$7) are located just outside the North and 
West borders of the CO gas map, therefore they are reasonably genuine members of VMR-D; the 
remaining sources (30\%) with the same colours, if not belonging to VMR-D, could represent 
star forming regions at larger distances.

In summary, we find a definitely many more young sources associated with the cloud, but
confirming that active star formation behind and/or in the close 
neighbourhood of our cloud is going on as well.\\
The relative percentages of sources attributed to different evolutionary stages (see
Table~\ref{tab:tab4}) can be compared with those of other well studied star forming regions.
Schmeja et al. (2005), in particular, have investigated number ratios of sources 
in different evolutionary classes in several star forming regions ($\rho$ Ophiuchi, Serpens,
Taurus, Chamaleon\,I, IC348) basing on data obtained before than the Spitzer advent. They find,
on average, that Class\,I sources are $\sim$ 1-10\%, while  Class\,II/III sources are about 80-95\%
of the total. 
With the advent of Spitzer, these percentages have increased slightly in favor of younger
protostars: by combining both IRAC and MIPS data, Reach et al. (2004) classified as Class\,I
11\%  of the sources in IC1396A and  Muzerolle et al. (2004) 20\% of the sources in  NGC7129.
Finally, in L1630 a ratio of 0.25 is found between Class\,I and Class\,II protostars (Muzerolle et
al. 2005). A comparison of our statistics with these is not straightforward both because of the
different classification adopted there (e.g. flat spectrum objects are not included) and the fact that
this paper uses MIPS data only. A more meaningful and direct comparison is with the works of Harvey
et al. (2007) and of R07, who performed a very similar analysis as the one we do here, but on 
the Serpens star forming region and on the IC348 and NGC1333 clouds in Perseus. The percentage
of Class\,I vs. Class\,II objects is 6\% vs. 63\% (Serpens), 
6\% vs. 85\% (IC348) and 7\% vs. 67\% (NGC1333), thus strongly in favoring the latter.    
In contrast, we find percentages in VMR-D of 23\% of Class\,I and 28\%
of Class\,II.

Two possible alternatives could explain such a difference: {\it i)} VMR-D is
significantly younger than either Perseus or Serpens. Such an hypothesis is supported by 
the age estimates of 1-2 Myr derived in the Perseus cloud (Palla \& Stahler 2000, R07) and of
2 Myr derived in Serpens (Djupvik et al. 2006) as compared with an age of 10$^5$-10$^6$ yr 
towards the clusters of VMR-D (Massi et al. 2000); {\it ii)} our K$_s$ vs. K$_s$-[24] 
diagram suffers from missing two important categories of sources. One category 
is represented by the $\sim$ 450 objects detected at 24$\mu$m, but without a
K$_s$-2MASS counterpart (see Table~\ref{tab:tab1}). The sensitivity limits of our
survey in terms of power density at a given wavelength are 
$\lambda$F$_{\lambda}$(2MASS) $\simeq$ $\lambda$F$_{\lambda}$(24-MIPS) $\sim$ 6 10$^{-16}$ W m$^{-2}$. 
This implies that these 450 sources are objects whose SED is rising with
wavelength and thus they could be additional young objects that tend even to increase the already anomalous 
percentage of Class~I sources. The second category, however, is represented 
by the about 5 10$^4$ 2MASS objects not having a MIPS 24$\mu$m counterpart.
Their SEDs are allowed to decrease with increasing wavelength, therefore,
although many of them could be foreground or background objects unrelated with
the VMR population, they undoubtly represent a potential reservoir of Class~II
and III objects. It should be sufficient that a very small fraction of them
($\sim$ 1-2\%) were genuine Class~II/III sources to reduce significantly the
relative excess of Class~I in VMR, and then to increase the apparent age of the
region. In this view the disagreement with other star forming regions 
could be reconciled, by considering that in those cases the lower
background level implied by their location outside the Galactic plane
has permitted to reach detection limits at 24$\mu$m up to an order of magnitude
fainter than in Vela, therefore allowing to trace the SED also for faint 
K$_s$ sources that decline going from the near- to the far-infrared. In any case, we expect
to provide a more certain answer to this issue in the next future, by means of
forthcoming IRAC images covering the relevant spectral bands at more adequate
sensitivity.

\subsection{[24] vs. [24]-[70]}
\begin{figure}
\includegraphics[angle=0,scale=.50]{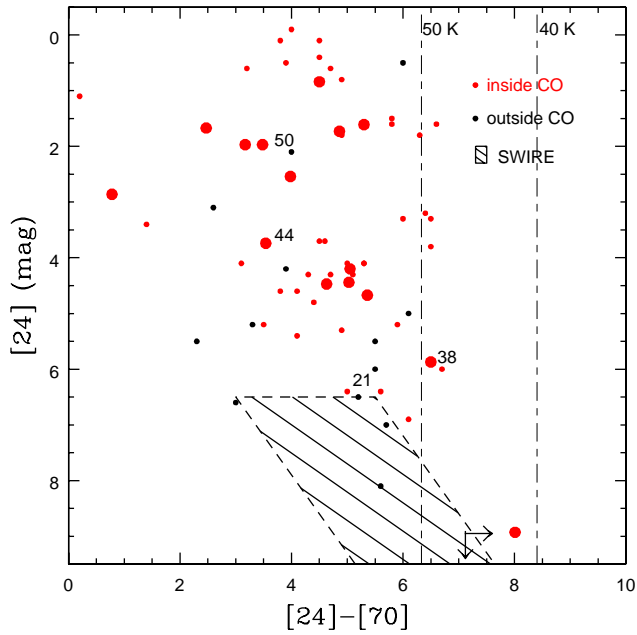}
\caption{Color-magnitude diagram [24] vs. [24]-[70], where only sources
not saturated at 24\,$\mu$m are plotted. Red/black dots refer
to sources inside/outside the CO contour map (41/12 sources). Large dots denote sources
associated with a dust core, while
numbered sources are the candidates exciting sources of the jets discussed
in Sect.\,6. The hatched area shows the {\it locus} of 
the SWIRE survey (taken from Fig.11 in R07).
\label{fig:24-70}}
\end{figure}

In Figure~\ref{fig:24-70} the color-magnitude diagram based on MIPS fluxes alone is
shown. Here the sources detected in both bands are plotted; the large majority
of them are on-cloud, although there is no clear difference between sources associated 
or not associated with dust cores (large dots). Remarkably, all sources (except 3) are located
to the right of [24]-[70]=2. This value pertains to SED's that increase with wavelength in such a
way that ($\lambda$F$_{\lambda}$)$_{70}$ = 2$\times$($\lambda$F$_{\lambda}$)$_{24}$.
These red objects are much more numerous than the 70~$\mu$m detections depicted in
Figure~\ref{fig:k-24}, since the majority of them lack a 2MASS counterpart.
Forthcoming IRAC data will help us to reconstruct their SED's more adequately, giving
constraints on their luminosity and evolutionary stage. A few sources (6) lie 
in the locus corresponding to black-body temperatures ranging between 40 and 50 K. These
are values theoretically predicted (Shu, Adams \& Lizano, 1987) for a collapsing isothermal
sphere, identified as Class 0 objects. These 6 sources are all located inside the CO cloud;  two
of them lie within a mm core and one (\#38) is also associated to a compact H$_2$ jet
(see Sect.\,6). Although at the moment a quantitative evaluation of their sub-mm vs.
bolometric luminosity cannot be given, they still represent the best candidates for 
members of the youngest population within the cloud.

\section{MIPS associations with H$_2$ protostellar jets}\label{sec:jets} 
MIPS offers a chance to identify, for the first time, very embedded compact exciting sources of
molecular (H$_2$) jets, found by D07, that have so far remained undetected even at the longest IRAS
wavelengths. We recall once again that our completeness limits are 
significantly higher than the sensitivity limits, so that by
scrutinizing our data-base for the weakest MIPS sources in selected areas, we have been able to
discover new objects down to 0.7 mJy at 24~$\mu$m.
We used this technique to search for the sources driving a number 
of molecular jets that were previously found by narrow band imaging centered at the H$_2$
(1-0)S(1) line (2.12$\mu$m, see D07 for details). 
Although our H$_2$ driving source survey is still incomplete, we here identify out some
interesting cases in the VMR-D cloud.

The results of the correlation between H$_2$ maps with our MIPS maps are given
in Table~\ref{tab:tab6}; here we list each jet,  the length and the 
dynamical time of the jet itself (having adopted d=700 pc); the total flux detected at 2.12 $\mu$m and the corresponding
H$_2$ luminosity; 
the identification  of the exciting source in our MIPS
catalogue; the coincidence with a
dust mm-peak,  and finally an estimate of the source bolometric luminosity
obtained by summing up all the contributions from the near-IR (if
any) to the 1.2\,mm flux, as derived by M07 SIMBA map.
We identify several different morphologies among our sources, including discovering the driving
sources for three of the jets (namely \# 2, 4, 5), objects that were not detected in either near-IR (K
band) or the far-IR (N-band, IRAS, MSX). 

\begin{figure}
\includegraphics[angle=0, width=14cm]{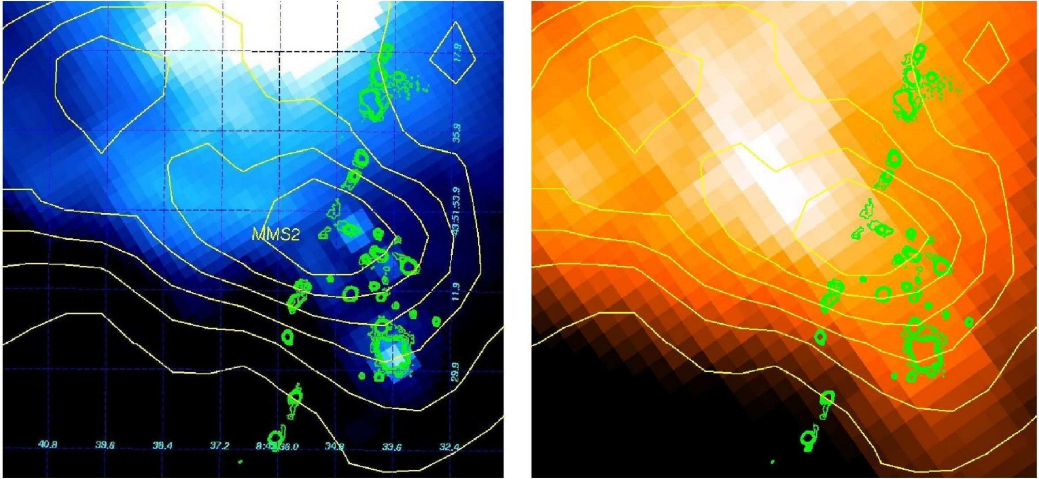}
\caption{H$_2$ contours of jet 1 (green) superposed on the MIPS 24~$\mu$m (left)
and 70~$\mu$m (right) images. Dust contours (from a 3$\sigma$ level in steps
of 3$\sigma$) are shown in yellow. Dust core MMS2 is located
at the jet center. \label{fig:jet1}}
\end{figure}

{\bf Jet 1 -} The jet center lies towards a millimeter peak (MMS2),
where no infrared source is detected down to K=17 mag. In the MIPS
24~$\mu$m band, an emission peak is found, although not aligned with the jet
axis. The lack of any aligned source suggests two possible alternative
scenarios: (i) we are observing just one jet lobe or (ii), more
reasonably, the exciting source is too faint to be detected even by
MIPS (F(24) $<$ 0.7 mJy). Scheduled APEX observations will hopefully
answer this question (Figure~\ref{fig:jet1}).

\begin{figure}
\includegraphics[angle=0, width=14cm]{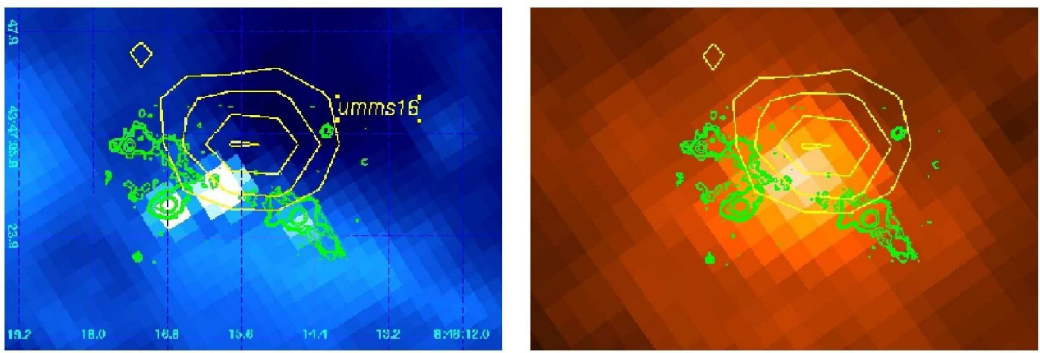}
\caption{The same as Fig.\ref{fig:jet1} for jet 2. Peak umms16 
(under-resolved at the SIMBA spatial resolution) is found near 
the jet center. The proposed exciting source is \#38 in 
Table~\ref{tab:tab3}.\label{fig:jet2}}
\end{figure}

\begin{figure}
\includegraphics[angle=0, width=14cm]{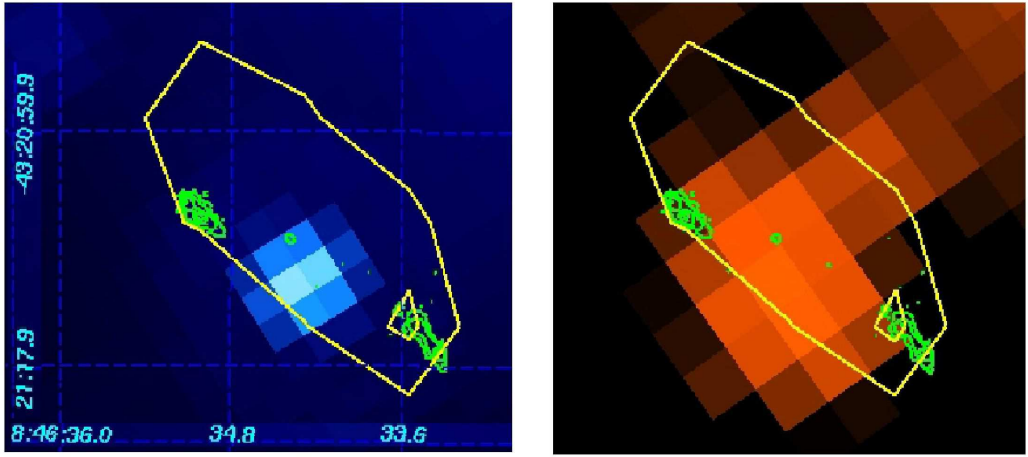}
\caption{The same as Fig.\ref{fig:jet1} for jet 3. A MIPS source (\#21)
is found at the jet center, corresponding to a location of weak dust emission. 
\label{fig:jet3}}
\end{figure}

{\bf Jets 2 and 3 -} Two small jets have been detected that correspond to faint dust peaks. The
exciting sources, although not detected in the infrared (NIR bands, IRAS, MSX), are clearly
recognizable in the MIPS images and one of them (\#38) is a
candidate Class 0 protostar. The compactness of these jets
implies a very short dynamical age, if reasonable conditions for
both shock velocity (v$_{shock}$ = 50 km s$^{-1}$) and inclination angle ({\it i} =
45$^{\circ}$) are assumed (Figures~\ref{fig:jet2} and \ref{fig:jet3}).

\begin{figure}
\includegraphics[angle=0, width=14cm]{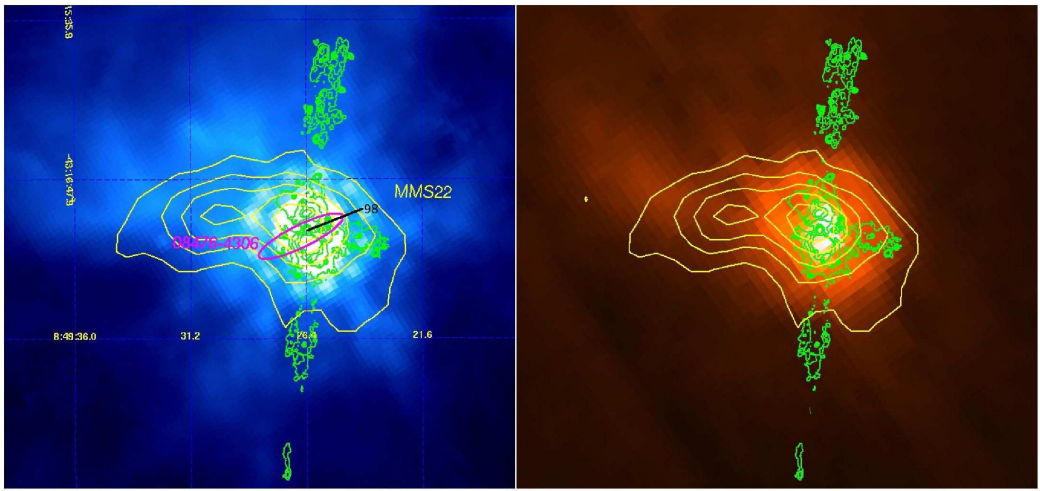}
\caption{The same as Fig.\ref{fig:jet1} for jet 4. Dust core MMS22 crosses
the jet. The candidate driving source (IRS20-MGL99\#98) is indicated.\label{fig:jet4}}
\end{figure}

{\bf Jet 4 -} A parsec scale jet emerges from a young near-infrared
cluster centered on IRAS08476-4306. The proposed exciting source,
detected in the near-IR bands is the IRS20-\#98
according to the Massi et al. (1999) list (Figure~\ref{fig:jet4}).

\begin{figure}
\includegraphics[angle=0, width=14cm]{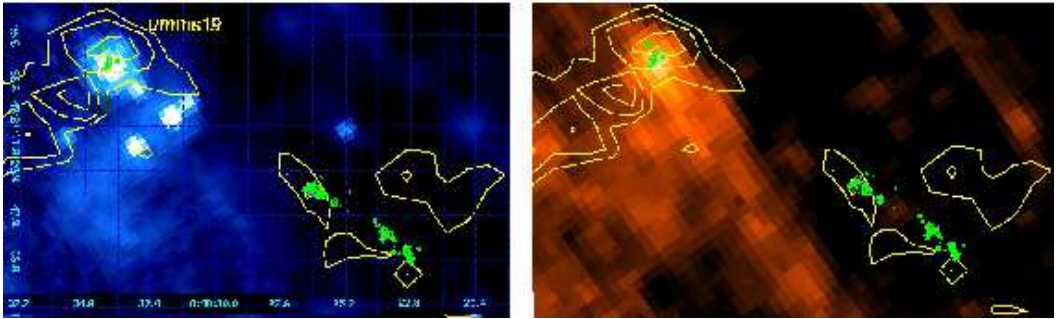}
\caption{The same as Fig.\ref{fig:jet1} for jet 5. The driving source is likely to be the 
MIPS source (\#44) associated with the under-resolved peak umms19. The lack of a
counter jet is probably due to the driving source being located near the edge 
of the H$_2$ image.\label{fig:jet5}}
\end{figure}

{\bf Jet 5 -} A point-like 24/70~$\mu$m source aligned with the jet
and corresponding to a dust peak  (umms19) is found about 2 arcmin
away towards the NE. If this source is indeed driving the
jet then we are observing just one jet lobe, being the
counter-jet located outside the H$_2$ investigated field. A NIR
cluster is also found at the MIPS source position (Figure~\ref{fig:jet5}).

\begin{figure}
\includegraphics[angle=0, width=7cm]{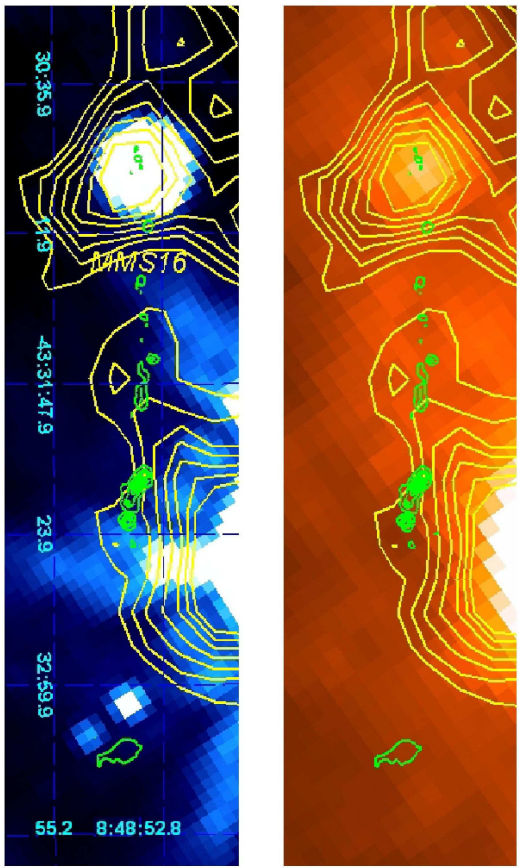}
\caption{The same as Fig.\ref{fig:jet1} for jet 6. The candidate exciting source is the 
24 and 70~$\mu$m source associated with mm peak MMS16 (\#50). Just one lobe 
has been detected. \label{fig:jet6}}
\end{figure}

{\bf Jet 6 -} A chain of H$_2$ knots emerges from a MIPS source (not
visible in the H$_2$ band) centered at the dust emission peak MMS16. 
We do not observe a counter jet (Figure~\ref{fig:jet6}). 

\clearpage

\section{Conclusions} 
MIPS maps covering 1.8 square degrees across the Vela Molecular Cloud D at 24 and
70\,$\mu$m are presented.  The data allowed us to derive the following results:
\begin{itemize}
\item[-] A total of 849 and 61 point sources at 24 and 70\,$\mu$m, respectively, have been
detected at completeness limits of 5 and 250 mJy.
\item[-] About half of the 24\,$\mu$m sources and two thirds of the 70\,$\mu$m ones 
are spatially located inside a region delimited by the $^{12}$CO contours (0.6 deg$^2$). 
The implication is that the IR source density doubles (and is four times when considering 
sources at 70\,$\mu$m) inside the CO contours as compared to outside the molecular cloud. A
quantitative analysis of the 24 and 70\,$\mu$m counts per deg$^2$ confirms this result. 
\item[-] The maps allow us to correlate MIPS sources with the distribution of the dust cores
found within VMR-D, and, when we extend the search of MIPS sources down to the
instrumental sensitivity limit, we find that most of these cores result associated with
red and cold objects.
\item[-] The MIPS sensitivity has enabled us to identify many new starless cores; the result will
prompt a revision of the relative percentages of young objects known so far in VMR-D. 
\item[-] IRAS-PSC detections of good quality (f$_{qual}$=3,2) are also seen by
MIPS, but only when the IRAS point-likeness confidence is high (correlation coefficient, cc,
equal to A or B). This result may be adopted as a broad confidence prescription 
for finding genuine point sources in the IRAS catalogue.
\item[-] About 400 MIPS sources have 2MASS K$_s$ counterparts.  Color-magnitude
plots constructed with magnitudes at 2.2, 24 and 70\,$\mu$m in VMR-D
show an excess of Class I objects in comparison with other well studied star formation 
regions.  This excess could be biased by the 
sensitivity limits of the 2MASS and MIPS surveys, otherwise, it
could reflect the short time elapsed since the first collapse of the cloud.
From the MIPS colors, 6 objects appear as potential candidates Class\,0 objects.
\item[-] We have detected the driving source in five out of six H$_2$ protostellar jets in VMR-D,
four of them embedded in mm-cores.  Such circumstance, along with the very low dynamical
time estimated for the jets, indicates ages of 10$^4$-10$^5$ yr for these sources.  
\item[-] We note that, given the southern location of VMR-D, many of the newly detected MIPS
sources will be excellent candidate targets for ALMA. 
\end{itemize}

\section{Acknowledgements}
This paper is based on observations
made with the Spitzer Space Telescope, which is operated by the Jet
Propulsion Laboratory, California Institute of Techonology under
a contract with NASA.  HAS acknowledges partial support from NASA grant NAG5-10659.

\begin{deluxetable}{cccc}
\tabletypesize{\scriptsize} \tablewidth{0pt}
\tablecaption{Statistics of MIPS point sources.\label{tab:tab1}} 
\tablehead{
item                 &   overall   &  inside CO contours$^a$   &  outside CO
contours$^b$            
}
\startdata
\cline{1-4}
24$\mu$m	      &  849       &  378          & 471 \\
70$\mu$m	      &   61       &   41          & 20 \\
24$\mu$m \& 70$\mu$m$^c$ &  52       &   40          & 12  \\
only 70$\mu$m	      &   9        &    1          & 8 \\
24$\mu$m \& 2MASS-K$_s$$^d$ &  401   &   180    & 221 \\
24$\mu$m \& 70$\mu$m \& 2MASS-K$_s$$^d$ & 28$^c$      &   23      &  5 \\
24$\mu$m \& dust peak$^e$ & 58      &  55    &     3      \\
70$\mu$m \& dust peak$^e$ & 19      &  19    &     0      \\ 
\cline{1-4}
\enddata
\tablenotetext{a}{size=0.61 deg$^2$; $^b$ size=1.23deg$^2$; $^c$coordinates coincident within
20$^{\prime\prime}$; $^d$coordinates coincident within 5$^{\prime\prime}$; $^e$coordinates coincident within 24$^{\prime\prime}$ 
(SIMBA HPBW). A total of 58 (19) sources at 24 (70)~$\mu$m result associated with 22 (resolved) and 20 (under-resolved) 
{\it mm} peaks, as listed by M07.Note that the area mapped in dust continuum
at 1.2 mm is slighthy larger than that covered with CO(1-0) observations (see Figs.~\ref{MIPS24},~\ref{MIPS70}).}
\end{deluxetable}



\begin{table*}
\begin{tiny}
\begin{center}
\caption{MIPS sources in the Vela Molecular Cloud-D. \label{tab:tab2}}
\begin{tabular}{cclcccccccc}
\tableline\tableline
Spitzer name & $\alpha$(2000.0)& $\delta$(2000.0)&
F$_{24}$ & $\Delta$F$_{24}$ & F$_{70}$ & $\Delta$F$_{70}$ &CO contours & dust core$^{\dag}$ \\
   &  {(h~~m~~s)} &   (${\circ}$~~${\prime}$~~${\prime\prime}$) &(mJy)  & (mJy)&(mJy)  & (mJy)&  &  \\
\tableline
SSTVMRD J084310.3-440052.7 & 8 43 10.3 & -44 00 52.7 & 4.5 & 0.2 & - & - & N & - \\        
SSTVMRD J084332.8-440301.0 & 8 43 32.8 & -44 03  01.0 & 31.6 & 0.2 & - & - & N & - \\        
SSTVMRD J084335.4-435539.0 & 8 43 35.4 & -43 55 39.0 & 8.2 & 0.1 & - & - & N & - \\        
SSTVMRD J084342.4-440134.5 & 8 43 42.4 & -44 01 34.5 & 2.3 & 0.1 & - & - & N & -  \\       
SSTVMRD J084347.4-435946.3 & 8 43 47.4 & -43 59 46.3 & 42.0 & 0.1 & - & - & N & - \\
\tableline
\end{tabular}
\end{center}
[The complete version of this table is in the electronic edition of
the Journal.  The printed edition contains only a sample.] \\
Notes to the table: $^{\dag}$: following the nomenclature by M07 and E07, dust peaks
are called MMS\# (umms\# for under-resolved peaks).
\end{tiny}
\end{table*}


\begin{sidewaystable*}
\begin{tiny}
\begin{center}
\caption{MIPS sources detected at 70$\mu$m. \label{tab:tab3}}
\begin{tabular}{cccccccccc}
\tableline\tableline
ID$^{\dag}$& Spitzer name & $\alpha$(2000.0)& $\delta$(2000.0)&($\Delta\alpha$/$\Delta\delta$)$_{70}$  &
(F$\pm\Delta$F)$_{24}$& (F$\pm\Delta$F)$_{70}$ & CO contours& dust core$^{\dag\dag}$ & ($\Delta\alpha$/$\Delta\delta$)$_{mm}$ \\
&   &  {(h~~m~~s)} &   (${\circ}$~~${\prime}$~~${\prime\prime}$)
   &(${\prime\prime}$/${\prime\prime}$) &(mJy)  & (mJy)&  & & $({\prime\prime}$/${\prime\prime}$) \\
\tableline
1$^a$ &	SSTVMRD J084409.5-440018.3	& 08~~44~~09.5 & -44~~00~~18.3    &  \nodata  &   $<$ 1.2      & 75$\pm$3      & N &  &\\
2   & SSTVMRD J084431.7-435005.3&		08~~44~~31.7& -43~~50~~05.3   &  4.5/4.5  & 4.3$\pm$0.1  &  78$\pm$3   &   N  &  &\\
3   & SSTVMRD J084444.6-434214.2&		08~~44~~44.6& -43~~42~~14.2   &  12.0/1.7  & 44.6$\pm$0.1  & 41$\pm$3   &   N  &  &\\
4   & SSTVMRD J084444.9-434927.4&		08~~44~~44.9& -43~~49~~27.4   &  6.0/0.3  & 16.6$\pm$0.1  &  28$\pm$3   &  N   & &\\
5    & SSTVMRD J084509.3-433802.7&		08~~45~~09.3& -43~~38~~02.7    &16.5/1.2 &   18.6$\pm$0.2 & 234$\pm$3    &  N & &\\  
6    & SSTVMRD J084533.3-434952.8&		08~~45~~33.3& -43~~49~~52.8    &  0.0/18.0&  3301$\pm$4    &  31882$\pm$263  & Y & MMS1&0.9/27.2\\ 
7$^c$ & SSTVMRD J084535.5-435107.2&	08~~45~~35.5& -43~~51~~07.2$^d$&  \nodata    &  $>$4000$^b$    & 22491$\pm$449 & Y & &\\ 
8$^e$ & SSTVMRD J084536.7-435155.4&	08~~45~~36.7& -43~~51~~55.4   &  0.0/ 16.5&  3097$\pm$6    &  34414$\pm$448  & Y &  &\\ 
9$^e$ & SSTVMRD J084537.0-435134.0&	08~~45~~37.0& -43~~51~~34.0   &  4.5/ 4.9 &  2740$\pm$6    & 34414$\pm$448  & Y &   &\\
10    & SSTVMRD J084541.1-435146.9&	08~~45~~41.1& -43~~51~~46.9   &  3.0/ 2.4 &  1611$\pm$69   & 36680$\pm$331  & Y  & MMS3 &15.3/14.9\\
11   & SSTVMRD J084544.0-432710.7&		08~~45~~44.0& -43~~27~~10.7   &  1.5/0.2  & 405.6$\pm$0.2  &  465$\pm$4   &  N   & &\\
12   & SSTVMRD J084546.7-432326.0&		08~~45~~46.7& -43~~23~~26.0   &  16.5/1.1  & 60.0$\pm$0.3  &  139$\pm$4   &   N  & &\\
13   & SSTVMRD J084604.4-433936.5&		08~~46~~04.4& -43~~39~~36.5   &  0.0/0.0   & 1041$\pm$4    &  4645$\pm$34   & N   & &\\
14$^a$ & SSTVMRD J084624.2-433415.6&	08~~46~~24.2  & -43~~34~~15.6 &  \nodata    &  diffuse      &   217$\pm$5    & N  &  &\\   
15$^e$ &SSTVMRD J084626.4-434247.5& 	08~~46~~26.4  & -43~~42~~47.5 & 18.0/ 3.2 &   245.6$\pm$0.8&  1712$\pm$5  & Y  & umms1 &9.4/20.9\\
16   & SSTVMRD J084626.7-434217.3&		08~~46~~26.7  & -43~~42~~17.3 &  6.0/ 6.7	&   141.4$\pm$0.3&  1612$\pm$7   & Y   & umms1 &13.9/9.3\\
17$^e$  & SSTVMRD J084627.3-434239.5&	08~~46~~27.3  & -43~~42~~39.5 &  4.5/ 4.8 &   234.6$\pm$0.2&  1712$\pm$5   & Y   &  umms1&22.9/12.9\\  
18   & 	SSTVMRD J084631.4-435605.7&	08~~46~~31.4  & -43~~56~~05.7 &  3.0/ 1.1 &  4366$\pm$10 	 &  16790$\pm$91 & Y   &   &\\  
19$^e$& SSTVMRD J084631.6-435532.2&	08~~46~~31.6  & -43~~55~~32.2 &  7.5/ 7.2 &   367$\pm$9    & 14939$\pm$68 & Y   &  &\\  
20$^e$& SSTVMRD J084633.1-435539.6&	08~~46~~33.1  & -43~~55~~39.6 & 15.0/ 0.2 &   349$\pm$11   &  14939$\pm$68 & Y   & &\\  
21   & SSTVMRD J084634.3-432115.1&		08~~46~~34.3  & -43~~21~~15.1 &  3.0/ 0.4	&    18.9$\pm$0.1&    211$\pm$6  & Y  &  & \\
22$^f$& SSTVMRD J084634.9-435437.1&	08~~46~~34.9  & -43~~54~~37.1$^d$ &  \nodata & $>$4000$^b$     & 78368$\pm$173  & Y   & MMS4  &5.1/1.1\\ 
23   & SSTVMRD J084637.4-435217.0&		08~~46~~37.4  & -43~~52~~17.0 &  1.5/ 6.7	&  4292$\pm$9 	 &  9024$\pm$45  & Y   &  &\\   
24   & SSTVMRD J084637.5-435357.1&	08~~46~~37.5  & -43~~53~~57.1 &  6.0/ 0.8	&  4758$\pm$8 	 &31250$\pm$180   & Y   &  &\\ 
25$^e$& SSTVMRD J084639.2-435254.1&	08~~46~~39.2  & -43~~52~~54.1 &  3.0/ 12.4&   341$\pm$5    &  8975$\pm$110& Y   &  &\\
26$^e$&	SSTVMRD J084639.5-435314.1&	 08~~46~~39.5  & -43~~53~~14.1 &  7.5/ 7.6 &   215$\pm$6	  &  8975$\pm$110& Y   & & \\ 
27   & SSTVMRD J084712.1-432250.6&		08~~47~~12.1  & -43~~22~~50.6 &  1.5/ 1.9 &   102.1$\pm$0.3&    360$\pm$6  &  Y  & & \\ 
28   & SSTVMRD J084714.4-431828.8&		08~~47~~14.4  & -43~~18~~28.8 &  7.5/ 1.5 &   314.5$\pm$0.1&    127$\pm$5  &  Y  & & \\
29   & SSTVMRD J084725.3-434900.1&		08~~47~~25.3  & -43~~49~~ 0.1 &  6.0/ 0.5 &   511.7$\pm$0.4&    114$\pm$6  &  Y  & & \\  
30   & SSTVMRD J084731.2-435357.0&		08~~47~~31.2  & -43~~53~~57.0 &  4.5/ 0.7 &    47.7$\pm$0.1&    229$\pm$4  &  Y  & & \\	   
31   & SSTVMRD J084731.7-434555.7&		08~~47~~31.7  & -43~~45~~55.7 &  7.5/ 4.4 &  2677$\pm$2	 &    359$\pm$10 &  Y  &  &\\   
32$^a$& SSTVMRD J084738.0-434255.3&		08~~47~~38.0  & -43~~42~~55.3 &   \nodata   &  diffuse       &	1653$\pm$16 &  Y &  &\\
33   & SSTVMRD J084742.8-434352.4&	08~~47~~42.8  & -43~~43~~52.4 & 13.5/ 1.7 &  1533.5$\pm$0.7&   1629$\pm$12 &  Y  & umms11 &4.8/13.4\\   
\tableline
\end{tabular}
\end{center}
\end{tiny}
\end{sidewaystable*}
\addtocounter{table}{-1}
\begin{sidewaystable*}
\begin{tiny} 
\begin{center}
\caption{MIPS sources detected at 70$\mu$m (\it{continued}).}
\begin{tabular}{cccccccccc}
\tableline\tableline
ID$^{\dag}$& Spitzer name & $\alpha$(2000.0)& $\delta$(2000.0)&($\Delta\alpha$/$\Delta\delta$)$_{70}$  &
(F$\pm\Delta$F)$_{24}$& (F$\pm\Delta$F)$_{70}$ & CO contours& dust core$^{\dag\dag}$ & ($\Delta\alpha$/$\Delta\delta$)$_{mm}$ \\
&   &  {(h~~m~~s)} &   (${\circ}$~~${\prime}$~~${\prime\prime}$)
   &(${\prime\prime}$/${\prime\prime}$) &(mJy)  & (mJy)&  & & $({\prime\prime}$/${\prime\prime}$) \\
\tableline
34   & SSTVMRD J084748.4-432536.4&		08~~47~~48.4  & -43~~25~~36.4 &  9.0/ 5.9 &   142$\pm$2    &  796$\pm$11 &  Y &umms12 & 24/19\\ 
35   & SSTVMRD J084751.7-432523.4&		08~~47~~51.7  & -43~~25~~23.4 & 10.5/13.5 &   137$\pm$3    &   1114$\pm$6  &  Y  &   &\\
36   & SSTVMRD J084755.7-441119.0&		08~~47~~55.7  & -44~~11~~19.0 & 15.0/ 1.9 &   53.6$\pm$0.6 &    518$\pm$5  &  Y  &  & \\
37   & SSTVMRD J084811.3-432056.0&		08~~48~~11.3  & -43~~20~~56.0 & 15.0/ 0.6 &    83.4$\pm$0.3&    504$\pm$6  &  Y  &   &\\
38$^e$& SSTVMRD J084815.8-434715.8&		08~~48~~15.8  & -43~~47~~15.8 &  1.5/ 0.1 &    32$\pm$1	 &   1410$\pm$7  &  Y  & umms16&1.2/8.0\\ 
39$^e$& SSTVMRD J084816.7-434719.4&		08~~48~~16.7  & -43~~47~~19.4 & 12.0/6.4  &    27$\pm$1   &   1410$\pm$7   &   Y & umms16&14.7/11.6\\
40   & SSTVMRD J084826.5-431721.2&		08~~48~~26.5  & -43~~17~~21.2 &  1.5/ 1.3 &    59.1$\pm$0.3&    162$\pm$6  & Y  &  &\\
41$^{a,g}$&SSTVMRD J084828.0-423630.8		&08~~48~~28.0  & -42~~36~~30.8 &   \nodata&   \nodata	      &	9538$\pm$49   &	 N & &\\
42$^{a,g}$&SSTVMRD J084829.3-423733.2		&08~~48~~29.3  & -42~~37~~33.2 &   \nodata&   \nodata	      &	3279$\pm$46   &  N &   &\\
43$^{a,g}$&SSTVMRD J084830.6-423559.9		&08~48~~30.6& -42  35  59.9    & \nodata  &   \nodata        & 12759$\pm$63  &  N &  & \\
44  &  SSTVMRD J084834.0-433051.3&	08~~48~~34.0  & -43~~30~~51.3  &  1.5/ 2.9 &   228$\pm$1 	 &  645$\pm$17   &  Y  & umms19/20&13.6/7.4;22.1/8.5 \\
45  & SSTVMRD J084841.6-433149.8&	08~~48~~41.6  & -43~~31~~49.8  &  1.5/ 2.0 &   116$\pm$2  	 &    903$\pm$7  &  Y  & MMS9  &20.4/1.8\\
46  & SSTVMRD J084844.2-431611.8&	08~~48~~44.2  & -43~~16~~11.8  &  6.0/ 0.3 &   100.0$\pm$0.2&    455$\pm$10 &  Y  &   &\\
47  & SSTVMRD J084846.4-425055.6&	08~~48~~46.4  & -42~~50~~55.6 & 12.0 13.6  &   45$\pm$1     &   776$\pm$10  &  N  & &\\
48$^h$&SSTVMRD J084848.2-425420.2&	08~~48~~48.2  & -42~~54~~20.2$^d$ &  \nodata  &  $>$4000$^b$    &125200$\pm$270  & N & &\\
49$^i$&SSTVMRD J084848.7-433230.7&	08~~48~~48.7  & -43~~32~~30.7$^d$ & \nodata   &  $>$4000$^b$    & 50640$\pm$43   & Y   & MMS12 &3.9/2.7 \\
50  & SSTVMRD J084853.2-433057.1&	08~~48~~53.2  & -43~~30~~57.1 &  0.0/ 2.6  &  1163$\pm$2    &   3126$\pm$30 & Y   & MMS16 &1.8/1.0\\
51  & SSTVMRD J084858.8-433825.1&	08~~48~~58.8  & -43~~38~~25.1 & 10.5/ 2.0  &   157$\pm$2 	 &   1758$\pm$18 & Y   & MMS17  &14.4/2.9\\
52  & SSTVMRD J084904.3-433805.0&	08~~49~~04.3  & -43~~38~~05.0 &  3.0/ 4.9  &   158$\pm$4    &  2359$\pm$28  & Y   & MMS18 &10.5/7.1\\
53  & SSTVMRD J084912.2-441636.5&	08~~49~~12.2  & -44~~16~~36.5 &  4.5/ 4.4  &    28$\pm$1    &  487$\pm$10   & N   &  &\\
54$^a$&	SSTVMRD J084912.5-432953.3	&08~~49~~12.5  & -43~~29~~53.3 &   \nodata    &    diffuse     &   650$\pm$10  &  N & & \\
55  & SSTVMRD J084913.1-433628.5&	08~~49~~13.1  & -43~~36~~28.5 &  0.0/ 0.9  &    685$\pm$1	 &  2910$\pm$20 & Y &MMS21&1.5/0.4\\ 
56  & SSTVMRD J084914.3-430019.8&	08~~49~~14.3  & -43~~00~~19.8 &  10.5/0.1  &   153.5$\pm$0.6  & 597$\pm$9   &  N   &&\\
57  & SSTVMRD J084915.3-433448.6&	08~~49~~15.3  & -43~~34~~48.6 & 16.5/13.3  &   158$\pm$2    &   2210$\pm$19 & Y   &  & \\
58  & SSTVMRD J084917.0-435600.3&	08~~49~~17.0  & -43~~56~~00.3 &  3.0/ 3.5  &  19.5$\pm$0.4	 &   375$\pm$6   & Y   & &\\  
59  & SSTVMRD J084921.2-440159.5&	08~~49~~21.2  & -44~~01~~59.5 &  1.5/ 6.7  &  158.3$\pm$0.1 & 295$\pm$5     & Y   &  &\\
60$^j$&	SSTVMRD J084926.2-431710.2&	08~~49~~26.2  & -43~~17~~10.2$^d$&  \nodata   & $>$4000$^b$     &38890$\pm$86    & Y  & MMS22&17.4/2.0\\
61  & SSTVMRD J084928.6-440429.2&	08~~49~~28.6  & -44~~04~~29.2 &  0.0/ 0.0  &    1.9$\pm$0.2 &  339$\pm$10   &  Y  & MMS23  &13.8/6.9\\ 
62$^k$&	SSTVMRD J084932.8-441050.0&	08~~49~~32.8  & -44~~10~~50.0$^d$&  \nodata   & $>$4000$^b$     & 44306$\pm$93   & Y   & MMS26 &15.6/9.9\\
63  & SSTVMRD J084936.1-441200.2&	08~~49~~36.1  & -44~~12~~0.2 &  7.5/ 2.5	&  1420$\pm$1	 &  13595$\pm$84  & Y  & MMS27  &23.1/4.2\\   
64  & SSTVMRD J084959.5-432300.7&	08~~49~~59.5  & -43~~23~~0.7 &  0.0/ 0.9  &   97$\pm$1     &   1472$\pm$13  &  Y  & umms26&6.9/5.1  \\
65$^a$ & SSTVMRD J085038.9-434948.8	&08~~50~~38.9  & -43~~49~~48.8 &  \nodata & diffuse  &  356$\pm$7   &  N   &&\\
66  &SSTVMRD J085129.7-433124.0&	08~~51~~29.7  & -43~~31~~24.0 & 16.5/ 2.2  &   74.1$\pm$0.2 &   2281$\pm$9   & N   &&  \\
67$^{a,g}$&SSTVMRD J085149.4-430540.2	&08~~51~~49.4& -43~~05~~40.2&   \nodata   &   \nodata        &   6656$\pm$33  &  N & &\\
\tableline
\end{tabular}
\end{center}
Notes to the table:$^{\dag}$: this ID is used for simplicity throughout the paper $^{\dag\dag}$:following the nomenclature by M07 and E07, dust peaks
are called MMS\# (umms\# for under-resolved peaks), while gas peaks are called 
VMR\#;$^a$:detected only at 70$\mu$m; $^b$ saturated; $^c$ IRS16 in
the Liseau et al. (1992) list . ;$^d$ 70$\mu$m coordinate; 
$^e$ associated with the same 70$\mu$m source;$^f$ IRS17; $^g$ source outside the 24$\mu$m map;
$^{h,i,j,k}$IRS18, IRS19, IRS20, IRS21 in the Liseau et al. (1992) list,
respectively.
\end{tiny}
\end{sidewaystable*}

\begin{deluxetable}{ccccccc}
\tabletypesize{\scriptsize} \tablewidth{0pt}

\tablecaption{IRAS PSC detections not recovered by MIPS\label{tab:tab4}} 
\tablehead{
PSC name    & MIPS 24$\mu$m & \multicolumn{2}{c|}{IRAS 25$\mu$m} & MIPS 70$\mu$m  &
\multicolumn{2}{c|}{IRAS 60$\mu$m} } \startdata                
            &                         &     fqual$^a$  &      CC$^b$               &                          &  fqual  &  CC  \\
       
\cline{1-7}                                                                                                                                                           
08441-4357  &  diffuse                &     1      & null                  &  missing                 &     3   & E     \\
08475-4255  &  diffuse                &     3      & D                     &  diffuse                 &     2   & D     \\
08475-4311  &  diffuse                &     3      & E                     &  diffuse                 &     2   & C     \\
08478-4303  &  intense knot           &     3      & A                     &  diffuse                 &     1   & null  \\
08479-4311  &  diffuse                &     3      & D                     &  diffuse                 &     1   & L     \\
08487-4250  &  diffuse (map edge)     &     1      & C                     &  diffuse                 &     3   & B     \\
08459-4338  &  diffuse                &     2      & D                     &  missing                 &     3   & C     \\
08489-4241  &  off edge               &     3      & C                     &  diffuse                 &     1   & D     \\
08457-4229  &  off edge               &     1      & E                     &  diffuse                 &     2   & D     \\
08460-4223  &  off edge               &     2      & D                     &  intense knot (map edge) &     2   & C     \\
08462-4235  &  diffuse (map edge)     &     1      & J                     &  diffuse?                &     3   & B     \\
08465-4230  &  off edge               &     2      & B                     &  diffuse? (map edge)     &     3   & C     \\
08471-4228  &  off edge               &     3      & G                     &  intense knot (map edge) &     2   & D     \\
08473-4235  &  off edge               &     3      & E                     &  diffuse (map edge)      &     1   & null  \\
08437-4323  &  diffuse                &     1      & D                     &  missing                 &     3   & B     \\
08468-4330  &  diffuse                &     3      & D                     &  missing                 &     2   & D     \\
08488-4308  &  diffuse                &     2      & C                     &  diffuse                 &     2   & E     \\
08490-4319  &  diffuse                &     1      & F                     &  missing                 &     3   & C     \\
08491-4310  &  diffuse                &     3      & B                     &  diffuse                 &     2   & C     \\
08491-4257  &  diffuse?               &     1      & E                     &  missing                 &     3   & B     \\
08493-4331  &  diffuse                &     1      & H                     &  missing                 &     3   & C     \\
08495-4306  &  diffuse                &     1      & C                     &  diffuse                 &     3   & C     \\
08477-4329  &  diffuse                &     1      & H                     &  diffuse                 &     3   & D     \\
08462-4400  &  diffuse                &     2      & D                     &  diffuse                 &     1   & D     \\
08463-4343  &  diffuse                &     2      & B                     &  diffuse                 &     3   & D     \\
08478-4403  &  diffuse                &     3      & C                     &  off edge                &     1   & J     \\
08478-4353  &  diffuse                &     1      & I                     &  diffuse                 &     3   & D     \\
\cline{1-7}
\enddata
\tablenotetext{a}{Flux density quality, encoded as 3: high quality, 2: moderate
quality, 1: upper limit; $^b$ point source correlation coefficient encoded 
as alphabetic character (A=100\%, B=99\%, ....N=87\%).}
\end{deluxetable}

\begin{deluxetable}{ccc}
\tabletypesize{\scriptsize} \tablewidth{0pt}
\tablecaption{Classification of the MIPS sources.\label{tab:tab5}} 
\tablehead{
item            & inside CO map &  outside CO map          
}
\startdata
photospheres    &  42(23\%)	  &  124(56\%)  \\
Class III	&  5 (3\%)	  &  11	(5\%)  	\\
Class II 	&  51(28\%)	  &  44	(20\%) 	\\
flat spectrum 	&  40(22\%)   	  &  31	(14\%) 	\\
Class I		&  42(23\%)	  &  11	(5\%) 	\\
Class 0		&  6	   	 &  1?$^a$      \\
Starless cores	&  5             &  0	      \\
\cline{1-3}
\enddata
\tablenotetext{a}{this source is observed only at 70$\mu$m, so that just upper
limits appear in the  24-[24-70] plot.}
\end{deluxetable}

\begin{deluxetable}{cccccccc}
\tabletypesize{\scriptsize} \tablewidth{0pt}
\tablecaption{Protostellar jets associated with MIPS sources in
VMR-D.\label{tab:tab6}} 
\tablehead{
Jet & length & T$_{dyn}$$^a$ & F$_{1-0S(1)}$& L$_{H_2}$$^b$ & Exc. source & 
  mm peak$^c$ & L$_{bol}$
}
\startdata
$\#$  & (pc)& (10$^3$ yr)  & (erg s$^{-1}$ cm$^{-2}$)& (10$^{-2}$L$_{\sun}$)& $\#$ & 
$\#$  & (L$_{\sun}$)\\
\cline{1-8}
  1    & 0.30 & 4.3 & 2.8 10$^{-13}$   &12  & -	      & MMS 2	 &  -	 \\
  2    & 0.13 & 1.7 & 1.5 10$^{-13}$   &6.6 & 38      & umms 16    & 0.3	 \\
  3    & 0.08 & 1.1 & 1.6 10$^{-14}$   &0.7 & 21      & 	-    	 & 0.07  \\
  4    & 0.68 & 93.5& 1.7 10$^{-12}$   &73  & 60      & MMS 22     & $>$14  \\
  5$^d$ & 0.18 & 5.0 & 7.6 10$^{-14}$   &3.4 & 44      & umms 19/20 & 0.5   \\
  6$^d$ & 0.49 & 15.3 & 8.5 10$^{-14}$  &3.7 & 50      & MMS16	  & 2.4   \\
\enddata
\tablenotetext{a}{computed for {\it i}=45$^{\circ}$ and v$_{shock}$=50 km s$^{-1}$; 
$^b$A$_V$=10 mag and L(H$_2$)=10$\times$L(2.12$\mu$m) are assumed; 
$^c$ names from M07; $^d$~just one lobe detected.}
\end{deluxetable}

\end{document}